\documentclass[letterpaper, 10 pt, conference]{ieeeconf}  

\IEEEoverridecommandlockouts                              

\usepackage{caption}
\usepackage{amssymb}  
\usepackage{multirow}

\usepackage{marvosym}
\usepackage{makecell}
\usepackage{amsmath}

\usepackage{graphics} 
\usepackage{epsfig} 
\usepackage{times} 
\usepackage{stfloats}
\usepackage{enumerate}
\usepackage{subcaption}
\usepackage{url}
\usepackage{longtable}
\usepackage{caption}
\usepackage{booktabs}

\title{\LARGE \bf
SurrealDriver: Designing LLM-powered Generative Driver Agent Framework based on Human Drivers' Driving-thinking Data
}

\author{Ye Jin, Ruoxuan Yang, Zhijie Yi, Xiaoxi Shen, Huiling Peng, Xiaoan Liu, Jingli Qin,\\ Jiayang Li, Jintao Xie, Peizhong Gao, Guyue Zhou and Jiangtao Gong\textsuperscript{\Letter}
\thanks{The authors are with the Institute for AI Industry Research, Tsinghua University, Beijing, China. Corresponding Email:
        {\tt\small gongjiangtao@air.tsinghua.edu.cn}}
}

\begin{document}

\maketitle
\thispagestyle{empty}
\pagestyle{empty}




\begin{abstract}

Leveraging advanced reasoning capabilities and extensive world knowledge of large language models (LLMs) to construct generative agents for solving complex real-world problems is a major trend. However, LLMs inherently lack embodiment as humans, resulting in suboptimal performance in many embodied decision-making tasks. In this paper, we introduce a framework for building human-like generative driving agents using post-driving self-report driving-thinking data from human drivers as both demonstration and feedback. To capture high-quality, natural language data from drivers, we conducted urban driving experiments, recording drivers' verbalized thoughts under various conditions to serve as chain-of-thought prompts and demonstration examples for the LLM-Agent. The framework's effectiveness was evaluated through simulations and human assessments. Results indicate that incorporating expert demonstration data significantly reduced collision rates by 81.04\% and increased human likeness by 50\% compared to a baseline LLM-based agent. Our study provides insights into using natural language-based human demonstration data for embodied tasks. The driving-thinking dataset is available at \url{https://github.com/AIR-DISCOVER/Driving-Thinking-Dataset}.


\end{abstract}

\section{INTRODUCTION}

Recently, remarkable advancements have been achieved in large language models (LLMs) known for their zero-shot prompting and common sense reasoning capabilities~\cite{yao2022react,liu2022few,brown2020language,hao2023reasoning,xi2023rise}. 
In addition to natural language tasks, LLMs, when equipped with specific sensory and control modules~\cite{huang2023embodied,chen20232}, can act as the decision-making core in executing embodied tasks, such as robotics and autonomous driving~\cite{chen2023towards,rajvanshi2023saynav,song2023llm}.
Previous research has validated the effectiveness of LLMs' advanced reasoning and extensive knowledge in embodied tasks~\cite{rajvanshi2023saynav,song2023llm}, but has also highlighted limitations in complex scenarios, like generating implausible sequences~\cite{wu2023embodied, wu2023plan} and a lack of operational experience~\cite{wang2023voyager}.
However, traditional demonstrations of embodied tasks are seldom suitable as examples for few-shot learning.
Current approaches primarily involve adjusting or constraining the LLM's task scope~\cite{wu2023plan} and enabling the LLM Agent to independently accumulate experience through environmental interactions~\cite{wang2023voyager,zhao2023expel}.

In the context of autonomous driving, agents analyze multimodal data, such as vectors~\cite{chen2023driving} and images~\cite{xu2023drivegpt4}, to make end-to-end driving decisions, demonstrated by projects like Driving with LLMs~\cite{chen2023driving} and DriveGPT4~\cite{xu2023drivegpt4}.
Compared to traditional fine-tuning, prompt-based methods with LLMs offer cost-effective and generalizable solutions~\cite{chen2023unleashing}. 
Approaches like Drive As You Speak~\cite{cui2024drive} and DiLu~\cite{wen2023dilu} integrate memory for coherent decision-making, and Drive Like a Human~\cite{fu2024drive} incorporate expert feedback to enhance performance.
However, these so-called human-like driving behaviors primarily rely on the human common sense inherent in LLMs.
LLMs acquire this common sense non-embodiedly from the noisy text corpus of the internet, lacking integration of professional, task-specific human data for embodied tasks~\cite{shanahan2023role}.
For LLM-based agents, employing human demonstrations~\cite{luo2023dr} and feedback~\cite{ouyang2022training} for reinforcement learning in embodied tasks such as driving proves prohibitively expensive.
A persistent challenge in this field is the lack of high-quality demonstrations and supervised human data.


To this end, in this paper, we innovatively leverage post-driving self-reports from human drivers, analyzing their thought processes as chain-of-thought prompts to enhance driving performance and human alignment in LLM-based agents. This approach offers new insights for aligning LLM-based agents with human drivers in embodied driving tasks.
We collected post-driving self-reports from 24 real-world drivers, detailing their considerations and decision-making processes during driving.
We then designed 'SurrealDriver,' an LLM-based framework for urban driving, grounded in four design considerations: a basic driving pipeline, a safety and memory mechanism, and human-aligned long-term driving guidelines, informed by demonstrations of human driving thought processes.
Our framework was evaluated through simulation experiments and human assessments, confirming its design efficiency.

Therefore, the contributions of this paper are as follows: 
\begin{itemize}
    \item The first high-quality human drivers' natural language-type driving-thinking dataset collected through an urban driving experiment;
    \item A generative driver agent framework 
 designed based on LLMs with human drivers' driving-thinking data as chain-of-thought prompts and implemented in Carla Simulator;
    \item An empirical validation of the effectiveness of our framework through simulation ablation experiments and human evaluation.
\end{itemize}

\section{Driving-thinking Dataset}
\subsection{Driving Experiment and data collections} 
To collect high-quality human drivers' language-type demonstration data, we invited 24 drivers (10 expert drivers and 14 novice drivers) to this driving-thinking Data collection session. Ten expert drivers were recruited through a formal career recruitment platform. They had extensive driving experience, ranging from 12 to 28 years, and their ages ranged from 35 to 48 years (M = 39.9, SD = 4.18). Novice drivers were recruited through social media, resulting in a group of 14 individuals aged between 20 and 25 years (M = 21.93, SD =1.49), with driving experience ranging from 1 to 4 years. This study was approved by the Institutional Review Board of the authors' institution. Before the experiment, all participants were ensured informed consent, acknowledging potential risks and their right to discontinue the study. To preserve participant confidentiality, all personal and confidential information has been anonymized, and the research results presented below have been subjected to de-identification.


To ensure the consistency between the collected natural language demonstrations and actual driving behaviors, we first had them participate in an actual complex urban road driving experiment and then we conducted post-driving interviews to collect their thinking-aloud data for safety reasons. For reviewing the driving experiment details in interview sessions, we recorded the driving process using multiple in-car cameras, including the driver's eye-tracking device (Tobii Glass 3\footnote{\url{https://www.tobii.com/products/eye-trackers/wearables/tobii-pro-glasses-3}}), roof-mounted 360-degree panoramic camera (Insta360 X3\footnote{\url{https://www.insta360.com/product/insta360-x3}}), and in-car motion camera (Dji OSMO Action 3\footnote{\url{https://store.dji.com/hk-en/product/osmo-action-3}}).
During the interviews, the drivers vocalized their decision-making process behind each driving behaviour as they reviewed the recorded footage. 
Besides, drivers were asked to contemplate the potential reasons behind their judgments and driving actions in complex driving scenarios during the experiment.


\subsection{Data Analysis and Dataset Construction}
Our data consists of 24 driver interview videos, with a duration ranging from 1.5 to 2 hours. We transcribed the audio recordings into written documents and organized the participants' descriptions of their driving decision processes for each scenario encountered during the experiments. Each participant's data was processed by two to three trained coders, and a coding consistency check was performed.

From our findings, an expert human driver doesn't just exhibit good driving behaviors by chance or intuition but continuously summarizes rules and patterns of driving behaviors. The construction of a thought chain progresses from strategic-level thinking to tactical-level decision-making and further to operational-level execution.
For example, most expert drivers reported that they observed different directions systematically while turning, no matter which direction they went in. As D11 (expert) shared,
\begin{quote}
D11 (expert): \textit{"No matter right or left, I must look at the direction that I turn to first because that's the road that I will take. However, I also look in the opposite direction. Basically, I look twice. The first time is to look at \underline{both sides}; the second time is \underline{to confirm}. Then I take the turns."} 
\end{quote}

Moreover, the expert drivers also had systematic, well-developed behavioral patterns when they interacted with other road users. For example, before entering the main road, the expert drivers evaluated the status of cars on the main road to decide when and how they got onto the main road.
\begin{quote}
D06 (expert): \textit{"Look at the left rearview mirror first, mainly about the \underline{speed of the back car}. If the speed is slow, I can step on gases and go directly. If the speed is fast, I can pause and wait. I can go after they pass by."} 
\end{quote}

We can see the thought chain of expert drivers is composed of multiple interconnected decision points, each based on the current traffic conditions and anticipated future changes. Such patterns not only enable human drivers to form muscle memory through repeated practice but can also be summarized into explicit chains of thought to teach autonomous driving algorithms based on LLMs.
Thus, we think that by using the driving-thinking data of expert drivers as prompts, these excellent driving behavior patterns can be expanded and generalized through LLMs. We compiled the "driving-thinking" data, along with demographic information and driving-related questionnaire data from the participants, into a dataset. This facilitates future research on driving behavior and the development of autonomous driving algorithms.

\begin{figure*}[h]
  \centering
  \includegraphics[width=\linewidth]{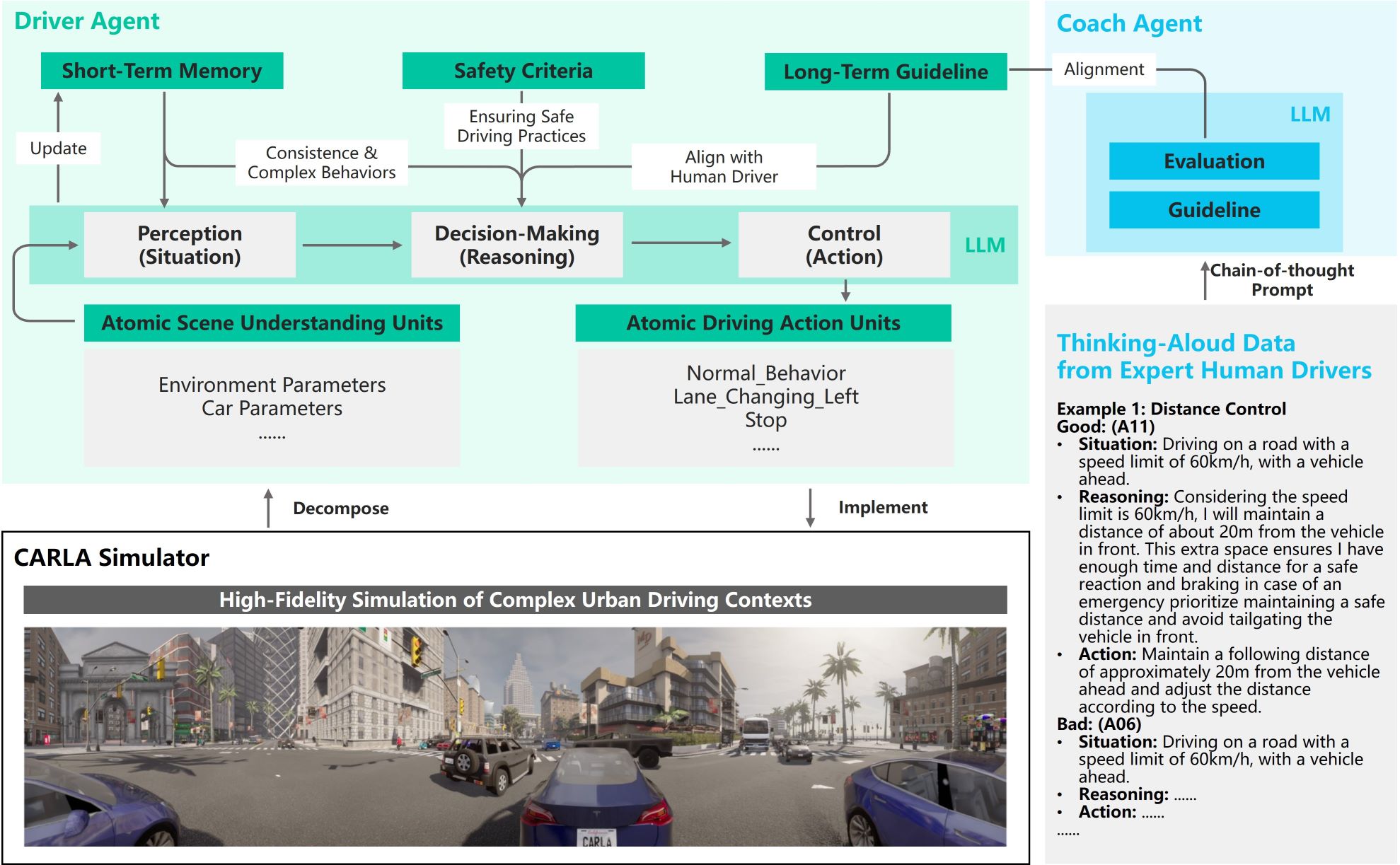}
  \caption{The framework of SurrealDriver.}
  \label{fig:SurrealDriver}
\end{figure*}

\section{SurrealDriver Framework}

\subsection{Framework Design}

Designing an agent capable of driving requires it to comprehend the complexity and diversity of driving environments, execute a continuous series of intricate operations, ensure safety, and harmonize with other human-driven vehicles. Based on these considerations, we have established the following framework as shown in Fig.~\ref{fig:SurrealDriver}:

\subsubsection{Perception: Atomic Scene and Atomic Actions.} Human driving scenarios are diverse, requiring agents to understand complex situations in detail. Traditional driving simulation methods train across a wide range of scenarios, which is costly. 
Our approach breaks down driving scenarios into discrete parameters for the LLMs. These parameters help the agent assess situations using common sense. We also simplify driving actions in the simulator into basic operations, enabling the agent to combine these for complex driving behaviors.

\subsubsection{Execution: Short-Term Driving Memory.} Effective car driving demands seamless and continuous actions, minimizing abrupt braking or sharp turns whenever feasible. Additionally, actions such as overtaking and following entail a fusion of fundamental maneuvers (e.g., acceleration, lane changing), rendering driving actions relatively intricate. 
To maintain smooth driving, we capture the agent's recent driving behavior over a few steps in the short-term driving memory module. These short-term driving memories aid the agent in sustaining consistency in decision-making. Moreover, the agent can employ these driving memories to amalgamate several basic driving operations for executing complex driving behaviors.
    
\subsubsection{Planning: Long-Term Human-like Driving Guidelines.} 
The agent must align its planning with that of human drivers. This module facilitates the agent in emulating the process by which humans learn from expert drivers to amass expertise and continually enhance their driving skills. 
To this end, we designed CoachAgent to assess the DriverAgent's driving behaviors and impart guidelines that must be adhered to. These guidelines are consistently integrated, contributing to the ongoing enhancement of the DriverAgent's driving proficiency.
    
\subsubsection{Overall Process: Strict Safety Criteria.} Ensuring safety is the most critical requirement for driving behavior simulation. Any simulated driving system must prioritize safety and establish rules within its framework to ensure the agent's safety.

Thus, throughout the entire driving process, safety should be consistently ensured through safety redundancy mechanisms. The agent is provided with stringent safety criteria to ensure the fundamental safety of the driving process.


\subsection{Implementation}
\begin{figure*}[h]
  
  \centering
  \includegraphics[width=0.9\linewidth]{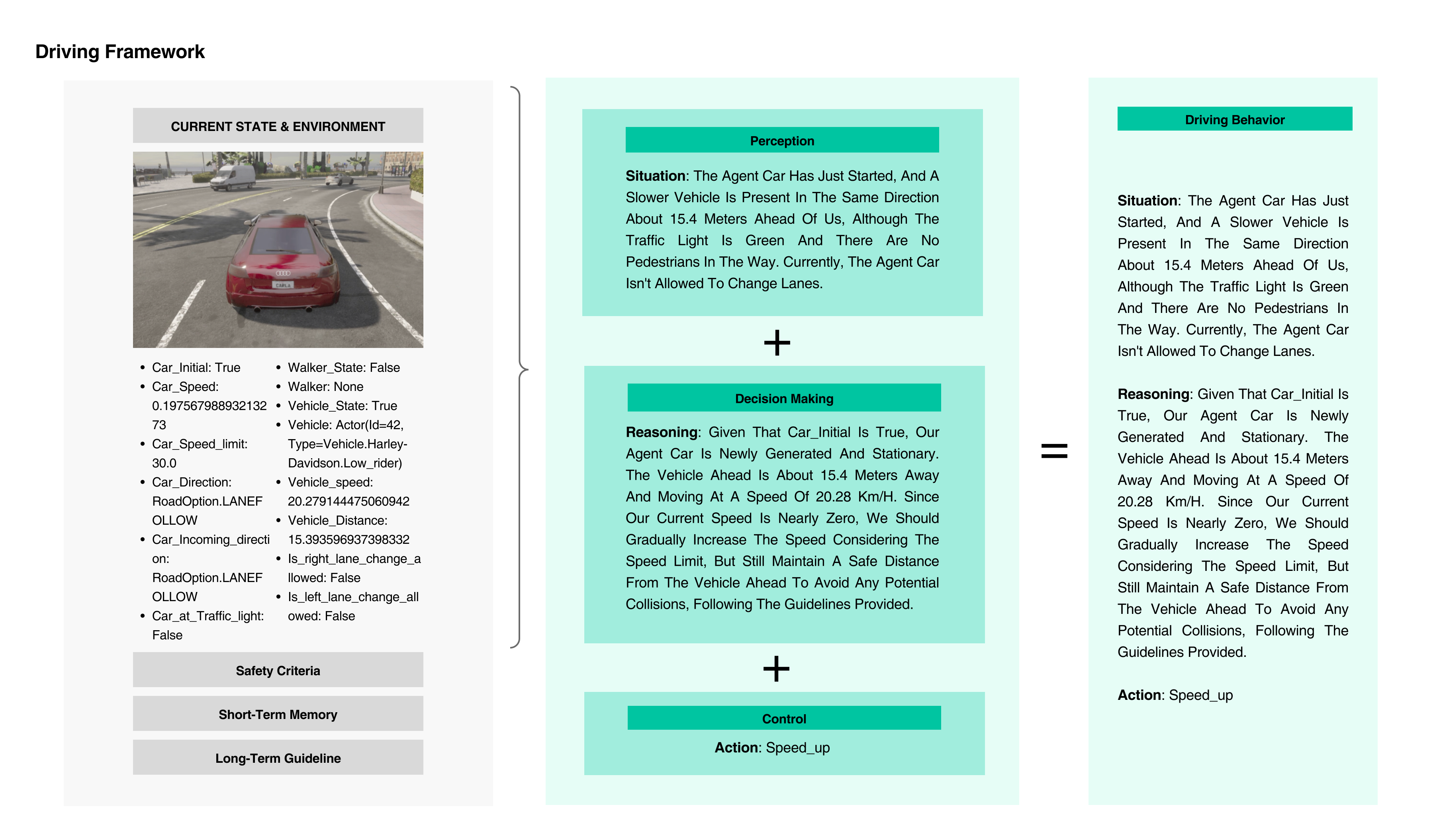}
  \caption{The Details of DriverAgent.}
  \label{fig:Driving_detail}
\end{figure*}

We built the SurrealDriver framework in the CARLA simulator~\cite{dosovitskiy2017CARLA}, including the basic driving pipeline, the memory and safety mechanism, and the human-aligned long-term driving guidelines.

\subsubsection{Basic Driving Pipeline.} As shown in Fig.~\ref{fig:Driving_detail}, the basic driving pipeline consists of three main processes: perception, decision-making, and control. 
In perception, DriverAgent receives and integrates vehicle and environmental data from the CARLA simulator. This data, provided as parameters, is analyzed based on predefined prompts and common sense, enabling DriverAgent to understand the vehicle's current situation.
Following perception, DriverAgent decides on the next steps, prioritizing safety and efficiency. It then proceeds to the control phase, where it sends JSON-formatted commands to CARLA, choosing from actions like stopping, maintaining speed, lane changing, or adjusting speed. These atomic actions allow DriverAgent to execute complex maneuvers based on the scenario.

\subsubsection{Memory and Safety Mechanisms}

The memory and safety mechanisms are built on top of the basic driving pipeline to store the information needed by the DriverAgent. It consists of three modules: Safety criteria and Short-term memory.
    
\textbf{Safety Criteria:} We implemented stringent safety criteria set to prevent hazardous maneuvers. The safety redundancy mechanism has two tiers. The first, mandatory tier, mandates actions like stopping if a vehicle or pedestrian is within 10 meters or at a red traffic light. The second, optional but recommended tier, includes decelerating when nearing vehicles or pedestrians within 20 meters, slowing down at intersections, keeping a minimum distance of 1 meter from moving cars, and optimizing energy use by reducing unnecessary speed changes.

\textbf{Short-term Memory:} To ensure the continuity and complexity of driving, we will store the driving behaviors of the current agent from the past few iterations and continuously update them, replacing the oldest with the latest to maintain a certain number of stored behaviors. These behaviors will then be provided to the DriverAgent again, becoming part of its perception.


\subsubsection{Human aligned Long-term Driving Guideline}\label{Section:CoachAgent}

\begin{figure*}[h]
  \centering
  \includegraphics[width=0.9\linewidth]{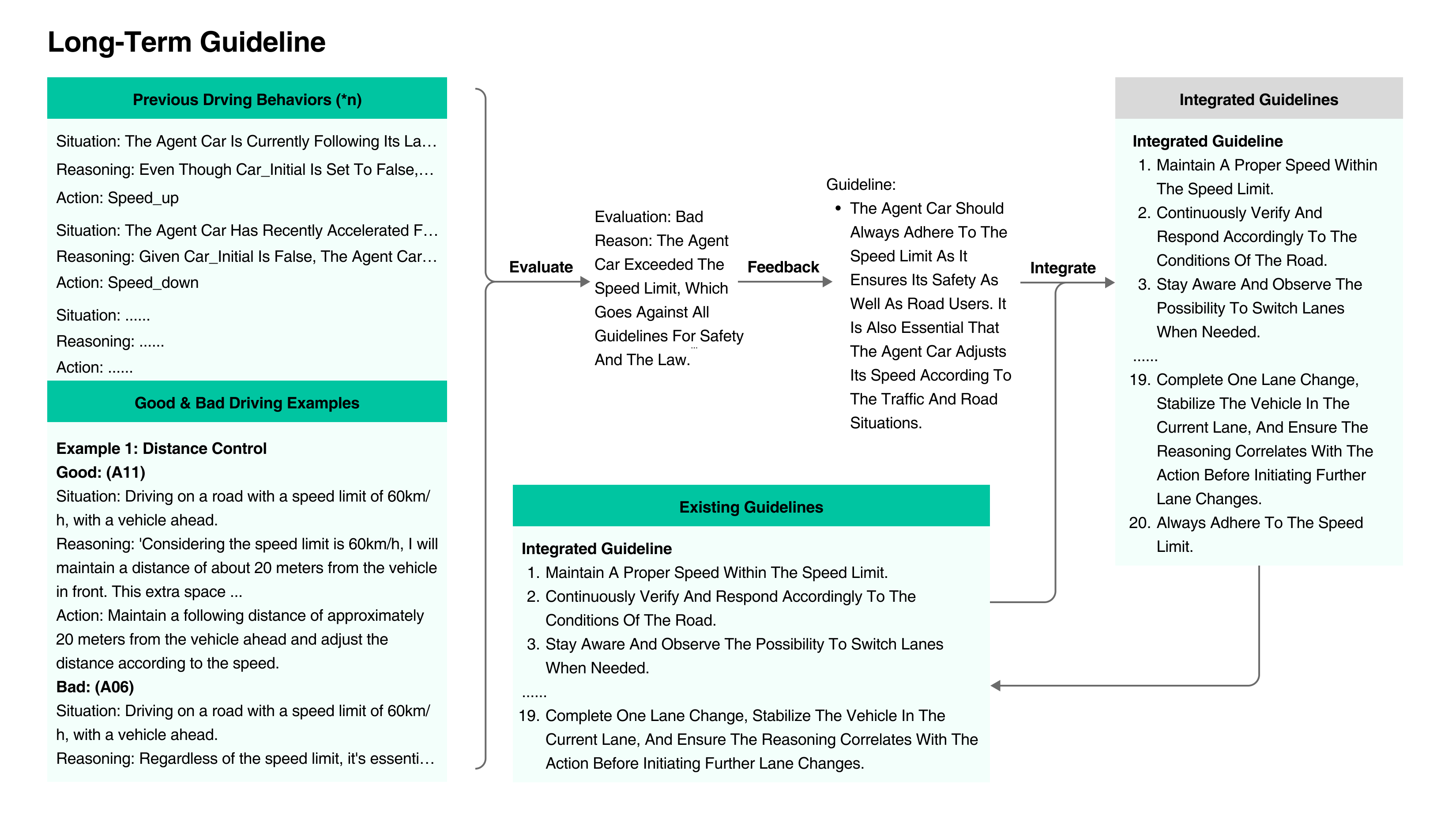}
  \caption{The CoachAgent for human alignment.}
  \label{fig:coachAgent}
\end{figure*}

To better align SurrealDriver with human drivers, we utilize the driving-thinking data of expert drivers collected in Section~\ref{Thnking-aloud} a chain-of-thought prompt. While designing examples, we followed a three-dimensional approach: situation, reasoning, and action as shown in Fig.~\ref{fig:coachAgent}. Situation provided specific road conditions during driver operations, and for each comparison case, we set the same road conditions, referencing the road conditions real drivers faced during their interviews. Reasoning was designed based on the content of driver interviews, with irrelevant information removed to make our examples concise and efficient in demonstrating human thinking and guiding the agent to learn human thought patterns.

\section{Evaluation}

We conducted driving experiments using agents from different frameworks in the same scenario, analyzing variations in their behaviors to understand how directives from different frameworks influence their driving. We evaluated the agents based on two primary dimensions: safety-driving capability and human-likeness. Safety-driving capability was assessed using an algorithmic experiment, while human-likeness was assessed through a human experiment.

\subsection{Algorithm Experiment}\label{AE}

\subsubsection{Experiment Environment Set-up}
The experimental setup on a ThundeRobot Zero desktop computer. The simulation environment was built upon the CARLA simulator version 0.9.14~\cite{dosovitskiy2017CARLA} and operated on Python 3.7 with Unreal Engine 4. The simulated environment was chosen to be Town10, and the Audi TT was the designated vehicle for all experiments, with fixed starting and continuously, randomly generated ending points for its path. Upon reaching the endpoint, another endpoint is randomly generated for continuous experiments. This process continues until the required number of driving rounds are completed. We leverage OpenAI's GPT-4 APIs for simulating drivers' driving decisions and solving related problems in a simulated environment. However, it takes several seconds for GPT-4 to make a decision, which is too long in a driving context for making immediate decisions. Therefore, we slowed down CARLA's simulation time based on the required token count by setting a fixed time step of 0.0006-0.0015 seconds.


\subsubsection{Results}
The overall experiment lasted 108405.90s (30.11 hours); the average experiment time for each condition was 7079.67s, 13730.6s, 23870.28s, and 63725.35s, respectively. We conducted statistical analyses separately for collision rates per unit distance and collision rates per unit time. The detailed results are shown in Table~\ref{tab:crach_freq}. Notably, we adjusted the algorithms controlling other vehicles and pedestrians to make them more prone to sudden maneuvers (e.g. abrupt lane changes, running red lights).
These edge cases aim to increase the risk level of the driving environment for the agent vehicle, making its driving performance more observable.

\begin{table}
  \caption{Collision Rate of Algorithm Experiment}
  \label{tab:crach_freq}
  \begin{tabular}{ccc}
    \toprule
    Framework &\begin{tabular}[c]{@{}l@{}}Collision Rate by\\ Distance (per meter)\end{tabular} &\begin{tabular}[c]{@{}l@{}}Collision Rate by\\Time (per second)\end{tabular}\\
    \midrule
\begin{tabular}[c]{@{}l@{}}w/o safety criteria,\\ w/o short-term memory, \\w/o long-term guidelines\end{tabular}  & 0.01453958
 & 0.041315485
 \\\hline
\begin{tabular}[c]{@{}l@{}}with safety criteria,\\ w/o short-term memory, \\w/o long-term guidelines\end{tabular} & 0.00923361
 & 0.02366976
 \\\hline
\begin{tabular}[c]{@{}l@{}}with safety criteria, \\with short-term memory, \\w/o long-term guidelines\end{tabular} & 0.005046864
 & 0.009530682
 \\\hline
Full framework & 0.002757353
 & 0.005100011
 \\

  \bottomrule
\end{tabular}
\end{table}

For the Safety Module, collision rate data shows that the framework with the safety module has a collision rate 57.46\% lower than the one without it. For example, in the absence of Safety Criteria, when the vehicle was at a distance of 5 meters from the preceding vehicle, the DriverAgent initiated a lane change, leading to a collision with the front vehicle. However, when running a framework with Safety Criteria, the vehicle encountered a situation where the distance to the preceding vehicle was 7 meters. Based on the information provided by the safety criteria, it initiated a stop behavior, safely coming to a halt behind the lead vehicle.

For the Short-term Memory Module, collision rate data shows that the framework with Short-term Memory has a collision rate 82.96\% lower than the one without it. We found that short-term memory plays an important role in enhancing the continuity of the agent's driving decisions. For example, in one experimental trial, the vehicle initially accelerated for a few steps, and when DriverAgent had to decide its next action, it had two options: to continue accelerating or to maintain its current speed. Considering its previous acceleration actions, it chose to maintain its current speed. 

For the Long-term Guidelines Module, collision rate data shows that the framework with Long-term Guidelines has a collision rate 83.03\% lower than the one without them. With long-term guidelines, the DriverAgent demonstrated an improvement in driving skills. For example, in one experimental trial, CoachAgent analyzed the initial driving behaviors and classified them as 'Bad.' The reason for this assessment was the excessive frequency of stopping. A guideline was generated that 'Maintain a consistent and safe speed.', which made the agent perform more human-like driving behaviour.

\subsection{Human Evaluation Experiment}

A single-factor within-subjects design was used to investigate how people rate each framework used in the algorithm experiment (see in Section~\ref{AE}).

\subsubsection{Experiment Design and Materials}
The independent variable was the framework, which included the “w/o safety, memory, or guideline framework” without safety criteria, short-term memory, or long-term guidelines; the “w/o memory or guideline framework” with safety criteria only; the “w/o guideline framework” with both safety criteria and short-term memory; and the “full framework” with safety criteria, short-term memory, and long-term guidelines. Therefore, the guideline framework was the full framework of SurrealDriver. The video of each framework was created by recording experiments in the algorithm experiment (see in Section~\ref{AE}). The length of each video is around 30 seconds.

\subsubsection{Participants and Procedures}
We invited another 24 adult participants (aged 29.3±4.9, male = 17, no overlap with participants in the Driving-thinking data collection experiment) with legal driving licenses to our human evaluation experiment. 
The experiment was conducted through online surveys. The survey started with demographic information questions including participants’ age, gender, phone number, driving silence status, years of driving experience, and kilometers of driving per month. Then the survey guided participants to watch videos embedded in the survey. All participants watched the videos in random order. After watching each video, they rate items that measure human likeness by asking whether the driver demonstrated driving operations like those conducted by human drivers using a 5-point Likert scale where 1 represented “not at all” and 5 presented “almost all.” 

\subsubsection{Results}
A one-way repeated measure ANOVAs were conducted to compare ratings among the four frameworks.
For human-likeness, the Huynh-Feldt correction was used because Mauchly’s test of sphericity was significant with epsilon values larger than 0.75. We found significant differences among the four frameworks: $F(2.5, 57.4) = 4.353$, $\textit{p} = 0.01$. The Bonferroni post hoc test revealed that the scores of the guideline framework were significantly higher than those of the w/o safety, memory, or guideline framework, $\textit{p} = 0.009$.

\section{Conclusion}

In our research, we developed SurrealDriver, an LLM-based driver agent framework. The results of both algorithm experiments and human evaluation indicate that this LLM-based driver agent framework offers better performance than the basic approach for driver simulations, bringing driver agent behavior closer to human-like driving and, consequently, simulating more realistic traffic environments.
By integrating human Driving-thinking data with LLMs, agents can utilize natural language and examples to add rules more conveniently, allowing for easier rule adjustments.

Thus, we provide the agent with the driving-thinking data of real drivers' behaviors obtained through interviews conducted during real vehicle experiments. The agent uses its capabilities based on LLMs to autonomously assess the quality of its driving behavior compared to detailed driving behaviour reasoning. It then enhances its driving skills based on the behavior of expert drivers. This approach differs from traditional reinforcement learning and other training methods by enabling the agent to learn directly from driver transcripts, similar to humans, without the need for translation into code. Our research provided valuable insights for future human-aligned agent generation.

\bibliographystyle{IEEEtran}
\bibliography{IEEEfull,root}

\begin{thebibliography}{10}
\providecommand{\url}[1]{#1}
\csname url@rmstyle\endcsname
\providecommand{\newblock}{\relax}
\providecommand{\bibinfo}[2]{#2}
\providecommand\BIBentrySTDinterwordspacing{\spaceskip=0pt\relax}
\providecommand\BIBentryALTinterwordstretchfactor{4}
\providecommand\BIBentryALTinterwordspacing{\spaceskip=\fontdimen2\font plus
\BIBentryALTinterwordstretchfactor\fontdimen3\font minus \fontdimen4\font\relax}
\providecommand\BIBforeignlanguage[2]{{%
\expandafter\ifx\csname l@#1\endcsname\relax
\typeout{** WARNING: IEEEtran.bst: No hyphenation pattern has been}%
\typeout{** loaded for the language `#1'. Using the pattern for}%
\typeout{** the default language instead.}%
\else
\language=\csname l@#1\endcsname
\fi
#2}}

\bibitem{yao2022react}
S.~Yao, J.~Zhao, D.~Yu, N.~Du, I.~Shafran, K.~Narasimhan, and Y.~Cao, ``React: Synergizing reasoning and acting in language models,'' \emph{arXiv preprint arXiv:2210.03629}, 2022.

\bibitem{liu2022few}
H.~Liu, D.~Tam, M.~Muqeeth, J.~Mohta, T.~Huang, M.~Bansal, and C.~A. Raffel, ``Few-shot parameter-efficient fine-tuning is better and cheaper than in-context learning,'' \emph{Advances in Neural Information Processing Systems}, vol.~35, pp. 1950--1965, 2022.

\bibitem{brown2020language}
T.~Brown, B.~Mann, N.~Ryder, M.~Subbiah, J.~D. Kaplan, P.~Dhariwal, A.~Neelakantan, P.~Shyam, G.~Sastry, A.~Askell, \emph{et~al.}, ``Language models are few-shot learners,'' \emph{Advances in neural information processing systems}, vol.~33, pp. 1877--1901, 2020.

\bibitem{hao2023reasoning}
S.~Hao, Y.~Gu, H.~Ma, J.~J. Hong, Z.~Wang, D.~Z. Wang, and Z.~Hu, ``Reasoning with language model is planning with world model,'' \emph{arXiv preprint arXiv:2305.14992}, 2023.

\bibitem{xi2023rise}
Z.~Xi, W.~Chen, X.~Guo, W.~He, Y.~Ding, B.~Hong, M.~Zhang, J.~Wang, S.~Jin, E.~Zhou, \emph{et~al.}, ``The rise and potential of large language model based agents: A survey,'' \emph{arXiv preprint arXiv:2309.07864}, 2023.

\bibitem{huang2023embodied}
J.~Huang, S.~Yong, X.~Ma, X.~Linghu, P.~Li, Y.~Wang, Q.~Li, S.-C. Zhu, B.~Jia, and S.~Huang, ``An embodied generalist agent in 3d world,'' \emph{arXiv preprint arXiv:2311.12871}, 2023.

\bibitem{chen20232}
P.~Chen, X.~Sun, H.~Zhi, R.~Zeng, T.~H. Li, G.~Liu, M.~Tan, and C.~Gan, ``\$ a\^{} 2\$ nav: Action-aware zero-shot robot navigation by exploiting vision-and-language ability of foundation models,'' \emph{arXiv preprint arXiv:2308.07997}, 2023.

\bibitem{chen2023towards}
L.~Chen, Y.~Zhang, S.~Ren, H.~Zhao, Z.~Cai, Y.~Wang, P.~Wang, T.~Liu, and B.~Chang, ``Towards end-to-end embodied decision making via multi-modal large language model: Explorations with gpt4-vision and beyond,'' \emph{arXiv preprint arXiv:2310.02071}, 2023.

\bibitem{rajvanshi2023saynav}
A.~Rajvanshi, K.~Sikka, X.~Lin, B.~Lee, H.-P. Chiu, and A.~Velasquez, ``Saynav: Grounding large language models for dynamic planning to navigation in new environments,'' \emph{arXiv preprint arXiv:2309.04077}, 2023.

\bibitem{song2023llm}
C.~H. Song, J.~Wu, C.~Washington, B.~M. Sadler, W.-L. Chao, and Y.~Su, ``Llm-planner: Few-shot grounded planning for embodied agents with large language models,'' in \emph{Proceedings of the IEEE/CVF International Conference on Computer Vision}, 2023, pp. 2998--3009.

\bibitem{wu2023embodied}
Z.~Wu, Z.~Wang, X.~Xu, J.~Lu, and H.~Yan, ``Embodied task planning with large language models,'' \emph{arXiv preprint arXiv:2307.01848}, 2023.

\bibitem{wu2023plan}
Y.~Wu, S.~Y. Min, Y.~Bisk, R.~Salakhutdinov, A.~Azaria, Y.~Li, T.~Mitchell, and S.~Prabhumoye, ``Plan, eliminate, and track--language models are good teachers for embodied agents,'' \emph{arXiv preprint arXiv:2305.02412}, 2023.

\bibitem{wang2023voyager}
G.~Wang, Y.~Xie, Y.~Jiang, A.~Mandlekar, C.~Xiao, Y.~Zhu, L.~Fan, and A.~Anandkumar, ``Voyager: An open-ended embodied agent with large language models,'' \emph{arXiv preprint arXiv:2305.16291}, 2023.

\bibitem{zhao2023expel}
A.~Zhao, D.~Huang, Q.~Xu, M.~Lin, Y.-J. Liu, and G.~Huang, ``Expel: Llm agents are experiential learners,'' 2023.

\bibitem{chen2023driving}
L.~Chen, O.~Sinavski, J.~H{\"u}nermann, A.~Karnsund, A.~J. Willmott, D.~Birch, D.~Maund, and J.~Shotton, ``Driving with llms: Fusing object-level vector modality for explainable autonomous driving,'' \emph{arXiv preprint arXiv:2310.01957}, 2023.

\bibitem{xu2023drivegpt4}
Z.~Xu, Y.~Zhang, E.~Xie, Z.~Zhao, Y.~Guo, K.~K. Wong, Z.~Li, and H.~Zhao, ``Drivegpt4: Interpretable end-to-end autonomous driving via large language model,'' \emph{arXiv preprint arXiv:2310.01412}, 2023.

\bibitem{chen2023unleashing}
B.~Chen, Z.~Zhang, N.~Langren{\'e}, and S.~Zhu, ``Unleashing the potential of prompt engineering in large language models: a comprehensive review,'' \emph{arXiv preprint arXiv:2310.14735}, 2023.

\bibitem{cui2024drive}
C.~Cui, Y.~Ma, X.~Cao, W.~Ye, and Z.~Wang, ``Drive as you speak: Enabling human-like interaction with large language models in autonomous vehicles,'' in \emph{Proceedings of the IEEE/CVF Winter Conference on Applications of Computer Vision}, 2024, pp. 902--909.

\bibitem{wen2023dilu}
L.~Wen, D.~Fu, X.~Li, X.~Cai, T.~Ma, P.~Cai, M.~Dou, B.~Shi, L.~He, and Y.~Qiao, ``Dilu: A knowledge-driven approach to autonomous driving with large language models,'' \emph{arXiv preprint arXiv:2309.16292}, 2023.

\bibitem{fu2024drive}
D.~Fu, X.~Li, L.~Wen, M.~Dou, P.~Cai, B.~Shi, and Y.~Qiao, ``Drive like a human: Rethinking autonomous driving with large language models,'' in \emph{Proceedings of the IEEE/CVF Winter Conference on Applications of Computer Vision}, 2024, pp. 910--919.

\bibitem{shanahan2023role}
M.~Shanahan, K.~McDonell, and L.~Reynolds, ``Role play with large language models,'' \emph{Nature}, vol. 623, no. 7987, pp. 493--498, 2023.

\bibitem{luo2023dr}
M.~Luo, X.~Xu, Z.~Dai, P.~Pasupat, M.~Kazemi, C.~Baral, V.~Imbrasaite, and V.~Y. Zhao, ``Dr. icl: Demonstration-retrieved in-context learning,'' \emph{arXiv preprint arXiv:2305.14128}, 2023.

\bibitem{ouyang2022training}
L.~Ouyang, J.~Wu, X.~Jiang, D.~Almeida, C.~Wainwright, P.~Mishkin, C.~Zhang, S.~Agarwal, K.~Slama, A.~Ray, \emph{et~al.}, ``Training language models to follow instructions with human feedback,'' \emph{Advances in neural information processing systems}, vol.~35, pp. 27\,730--27\,744, 2022.

\bibitem{dosovitskiy2017CARLA}
A.~Dosovitskiy, G.~Ros, F.~Codevilla, A.~Lopez, and V.~Koltun, ``Carla: An open urban driving simulator,'' in \emph{Conference on robot learning}.\hskip 1em plus 0.5em minus 0.4em\relax PMLR, 2017, pp. 1--16.

\end{thebibliography}

\end{document}